 \documentclass[final,3p,times,twocolumn]{elsarticle}

 \usepackage{graphicx}

\usepackage{amssymb}

\journal{Nuclear Physics B}

\begin{document}

\begin{frontmatter}


\title{Roles of Antiferromagnetic Fluctuation in High Field Phase Diagram of Superconductors with Strong Paramagnetic Depairing}


\author{Ryusuke Ikeda and Kazushi Aoyama}

\address{Department of Physics, Graduate School of Science, Kyoto University, Kyoto 606-8502, Japan}

\begin{abstract}
The high field phase diagram and magnetic properties of CeCoIn$_5$ below $H_{c2}(0)$ are examined from the picture regarding the high field and low temperature (HFLT) phase as a possible Fulde-Ferrell-Larkin-Ovchinnikov (FFLO) vortex lattice. Crucial roles of antiferromagnetic (AFM) fluctuations enhanced close to $H_{c2}(0)$ are stressed. The FFLO vortex lattice with a longitudinal modulation parallel to the field is stabilized compared with those with lateral modulations as a consequence of the presence of AFM fluctuations. Further, an unusual field-induced enhancement of the flux distribution is argued to be a consequence of interplay between the paramagnetic depairing and AFM fluctuation, both of which are enhanced with increasing field. 
\end{abstract}


\end{frontmatter}


\section{Introduction}
At present, it is believed that the high field and low temperature (HFLT) phase of the heavy fermion superconductor CeCoIn5 with a $d$-wave pairing symmetry is a Fulde-Ferrell-Larkin-Ovchinnikov (FFLO) vortex lattice with a paramagnetically induced modulation {\it parallel} to the applied magnetic field \cite{B1,RI1}. This identification is based on several key observations \cite{W,M,T} and their consistency with theoretical results \cite{RI1}.  For instance, the instability of HFLT phase via Cd or Hg doping \cite{T}, which induces a localized \cite{P} antiferromagnetic (AFM) order, seems to imply a destruction of long range order of the FFLO modulation and is well understood if the HFLT phase includes an {\it inhomogeneous} AFM order \cite{RI1}. In this sense, the FFLO picture of the HFLT phase of CeCoIn$_5$ is {\it compatible} with the recent observation of AFM order there \cite{K}. Nevertheless, a relation between the HFLT phase and the AFM {\it fluctuation}, which is believed to be enhanced near $H_{c2}(0)$ \cite{R}, has not been well understood yet. Here, we will report on some results of our study on the vortex state taking account of both the strong paramagnetic depairing {\it and} the AFM quantum fluctuation enhanced near $H_{c2}(0)$. 

\section{Phase diagram in ${\bf H} \perp c$}

As mentioned earlier \cite{RI2}, stability of the FFLO state with a modulation parallel to the field against those with lateral modulations is ensured by going {\it beyond} the weak-coupling BCS approach and including some quasiparticle damping. In CeCoIn$_5$ with strong AFM fluctuation {\it near} $H_{c2}(0)$, this fluctuation-induced damping seems to be a main origin of making the states with {\it lateral} modulations relatively unstable, although the AFM fluctuation tends to suppress even the FFLO state with the longitudinal modulation (see Fig.1). On the other hand, the AFM fluctuation does not necessarily suppress the first order $H_{c2}$-transition, which is another peculiar feature in CeCoIn$_5$ in high fields \cite{B1,W}, and seems to also have a partial role of inducing the first order $H_{c2}$ transition. An example of the $H$-$T$ phase diagram in ${\bf H} \perp c$ we obtain in terms of a microscopically-derived Ginzburg-Landau (GL) free energy functional including the AFM fluctuation is shown in Fig.1. Since an applied pressure in real experiments corresponds to a reduction of AFM fluctuations, the dependences of the two (mean-field) transition curves and the onset of the first order $H_{c2}$-transition on the AFM fluctuation strength are consistent with the features found in experiments \cite{M}. 

\begin{figure}[t]
\begin{center}
\includegraphics[scale=0.43]{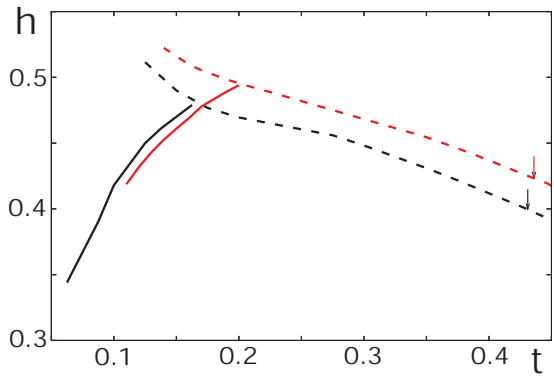}
\caption{The resulting $H_{c2}(T)$ (dashed curves) and the FFLO transition curves (solid ones) in ${\bf H} \perp c$ for a smaller (red) and larger (black) strength of the AFM fluctuation. Each arrow indicates the onset of the first order $H_{c2}$-transition. The upturn of $H_{c2}(T)$ below $t=0.2$ is an artifact of the expansion on the wavenumber of the FFLO modulation and can be improved. Here, $h$ is the reduced magnetic field normalized by the orbital-limiting field in 2D case, and $t=T/T_{c0}$. 
. \label{fig:ph}}
\end{center}
\end{figure} 

\section{Flux distribution in ${\bf H} \parallel c$}

Neutron scattering data on CeCoIn$_5$ in ${\bf H} \parallel c$ show \cite{B2} the vortex lattice form factor enhanced with increasing the field, which is an opposite trend to the conventional one in type II superconductors. Although such an anomalous behavior has been explained in intermediate fields as a result of strong Pauli-paramagnetic effect \cite{IM}, this interpretation is insufficient in higher fields where the AFM fluctuation is not negligible. If the damping effect suppressing the paramagnetic depairing is the main consequence of the AFM fluctuation, it seems difficult that the above-mentioned anomaly is attributed to the AFM fluctuation. However, the AFM fluctuation also leads to an effective increase of the electron correlation and thus, of the paramagnetic depairing, implying that the field dependence of roles of the AFM fluctuation is not intuitively clear. For this reason, we have investigated the flux density distribution of the vortex lattice based on the microscopically derived GL free energy. The flux distribution follows from the Maxwell equation. Expanding up to the lowest order in the squared pair field, $|\Delta|^2$, and also in a local value of AFM fluctuation correlator $\chi$ ($> 0$), we obtain the superconducting contribution to the longitudinal flux density parallel to the applied field in the form

\begin{equation}
B^{(s)}({\bf r}) = 4 \pi C_{\rm G} \, |\Delta|^2 \, [ \, \tilde{B}({\bf r}) + \chi \, \tilde{B}^{\rm AF}({\bf r}) \, ],
\end{equation}
where $C_{\rm G}$ is a scale factor measuring the magnitude of the local flux density in the familiar Abrikosov lattice near $T_c$. Figure 2 shows spacial dependences of $\tilde{B}({\bf r})$ and $\tilde{B}^{\rm AF}({\bf r})$ on the coordinate between neighboring two vortices for the two cases of a small (dotted lines) and large (solid lines) Maki parameter $\alpha_M$, which measures the Pauli-paramagnetic effect. The feature in $\tilde{B}$ that the magnetic flux is accumurated near the vortex center with increasing the paramagnetic depairing is qualitatively the same as that in Ref.\cite{IM}. A crucial result in this study is that, as the red curves show, the fluctuation contribution $\tilde{B}^{\rm AF}$ is competitive with $\tilde{B}$ for small Maki parameters, while it assists and {\it enhances} the flux accumuration in the vortex core induced by the paramagnetic effect for large enough $\alpha_M$. It appears that, among roles of the AFM fluctuation, the mass enhancement of quasiparticles will overcome an increase of the quasiparticle damping. This enhanced flux accumuration due to the AFM fluctuation should be a direct origin of the unresolved enhancement \cite{B2} of the structure factor in higher fields and seems to imply an apparent presence of an AFM quantum critical point near $H_{c2}(0)$. 

\begin{figure}[t]
\begin{center}
\includegraphics[scale=0.6]{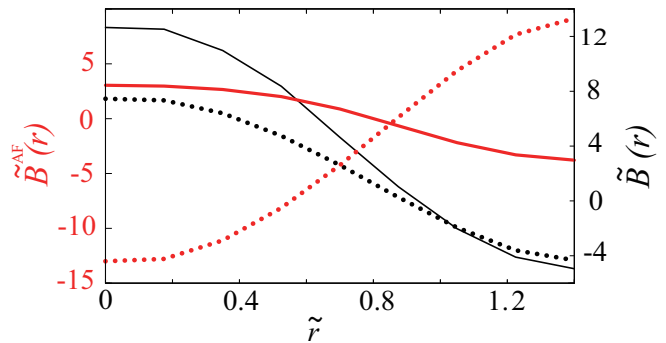}
\caption{Dependences of ${\tilde B}({\tilde {\bf r}})$ (black) and ${\tilde B}^{\rm AF}({\tilde {\bf r}})$ (red) at $t=0.1$ and $h=0.1$ on the coordinates ranging from a vortex center (${\tilde r}=0$) to the midpoint (${\tilde r}=1.4$) between the two vortices. Here, a triangular lattice was assumed in a d-wave pairing. The dotted and solid curves correspond to those for $\alpha_{M}=0.01$ and $6.5$, respectively. \label{fig:vortexcore}}
\end{center}
\end{figure} 

\section{Acknowledgement}
The work is supported by the Japan Society for the Promotion of Science.

\end{document}